\documentclass{sigchi-ext}

\usepackage[T1]{fontenc}
\usepackage{textcomp}
\usepackage[scaled=.92]{helvet} 
\usepackage{graphicx} 
\usepackage{balance}  
\usepackage{booktabs} 
\usepackage{ccicons}  
\usepackage{ragged2e} 

\usepackage{amsmath}
\usepackage{graphics}      
\usepackage[T1]{fontenc}   
\usepackage{txfonts}
\usepackage{mathptmx}
\usepackage{color}
\usepackage{textcomp}
\usepackage{subfig}

\newcommand{\hir}{$hir$ }
\newcommand{\hirWoS}{$hir$}

\newcommand{\mir}{$mir$ }
\newcommand{\mirWoS}{$mir$}

\newcommand{\lir}{$lir$ }
\newcommand{\lirWoS}{$lir$}

\def\plaintitle{Energy-Efficient Thermostats for Room-Level Air Conditioning}

\def\emptyauthor{}
\def\plainkeywords{Smart thermostat; Comfort; Feedback; Fault Detection}

\title{\plaintitle}

\numberofauthors{1}
\author{%
  \alignauthor{%
  	\textbf{Milan Jain}\\
    \affaddr{Indraprastha Institute of Information Technology Delhi}\\
    \email{milanj@iiitd.ac.in}}\\
}

\copyrightinfo{Copyright \textcopyright~2018 for this paper held by its author(s). Copying permitted for private and academic purposes.}

\definecolor{linkColor}{RGB}{6,125,233}
\hypersetup{%
  pdftitle={\plaintitle},
  pdfauthor={\emptyauthor},
  pdfkeywords={\plainkeywords},
  bookmarksnumbered,
  pdfstartview={FitH},
  colorlinks,
  citecolor=black,
  filecolor=black,
  linkcolor=black,
  urlcolor=linkColor,
  breaklinks=true,
}

\begin{document}

\maketitle

\RaggedRight{} 

\begin{abstract}
Room-level air conditioners (also referred as ACs) consume a significant proportion of total energy in residential and small-scale commercial buildings. In a typical AC, occupants specify their comfort requirements by manually setting the desired temperature on the thermostat. Though commercial thermostats (such as Tado) provide basic energy-saving features, they neither consider the influence of external factors (such as weather) to set the thermostat temperature nor offer advanced features such as monitoring the fitness level of AC. In this paper, we discuss grey-box modeling techniques to enhance existing thermostats for energy-efficient control of the ACs and provide actionable and corrective feedback to the users. Our study indicates that the enhancements can reduce occupants' discomfort by 23\% when maximising the user experience, and reduce AC energy consumption by 26\% during the power-saving mode. 
\end{abstract}


\section{Introduction}
	\label{sec:introduction}
Efficient and optimised usage of heating and air conditioning devices - a major power consuming appliance, can save significant energy across residential and commercial buildings~\cite{davis2015contribution}. Unlike big commercial buildings, people prefer window and split air conditioners (ACs) in residential and small-scale commercial which comes with an inbuilt thermostat for the occupants to mention their comfort requirement in the form of thermostat temperature. When thermostat embedded within the AC senses that room temperature is close to a lower threshold, the thermostat shuts down the compressor - a major power consuming component of the AC. Correspondingly, when the room temperature attains the upper threshold, the thermostat again turns on the compressor. Even though the control logic of thermostat drives the AC energy consumption and maintains user comfort, the AC manufacturers design the thresholds based on lab tests performed with numerous assumptions about the room and its thermal environment. 

In addition to that, the inbuilt thermsotats - 
\begin{enumerate}
	\item \textit{need manual intervention} - they rely on tenants to set the thermostat temperature.
	\item \textit{are unaware about the future} - the thermostats are incapable of leveraging the readily available information such as weather forecast.
	\item \textit{are unaware about the user comfort} - as the thermostats regulate the room temperature based on lower and upper thresholds, there is no way to monitor the impact of set temperature on user comfort.
	\item \textit{neglect the influence of weather conditions} - The thermostats assume a highly insulated room environment with negligble influence of weather conditions, however, the room temperature gets affected by the change in weather conditions.
	\item \textit{are incapable of monitoring the AC fitness level} - undetected faults usually make an appliance irreparable and useless. A smart thermostat capable of monitoring the fitness level of AC can help user in avoiding such instances.
\end{enumerate}

The existing commercial thermostats~(\cite{Sensibo}, \cite{Tado}) allow users to operate their AC locally as well as remotely (from any other location) through their smartphones, to reduce manual internvention. Additionally, the commercial thermostats can also pre-cool the space by monitoring the GPS location of the user. However, even the commercial thermostats fail to address many issues (of an inbuilt thermostat), as stated above. To further understand the perception of people towards the existing thermostats, we interacted with few residents and as per a user,

\emph{``Ahmm, setting a thermostat temperature is quite challenging task. I come home, feeling very hot and start the AC at 23... then in some time it becomes very cold and I turn it off... and after some time, turn it on again (may be at a higher temperature). You never know what to set AC on.... therefore, during night I fix my AC at 24-25 and use blanket... at some point I feel cold and take the blanket, after few minutes I start feeling hot and remove the blanket. Though, it's annoying but that's the best way I have figured out till now..."}

In this study, we explored grey-box modelling techniques to enhance smart thermostats for energy-efficient control while ensuring user comfort, and also provide actionable and corrective feedback to the occupants for optimal AC usage. First, we proposed a comfort-energy knob (for existing commercial thermostats) to maximise energy savings while ensuring the user comfort. Our analysis indicates that such a knob can reduce residents' discomfort by 23\% when optimising comfort, and save 26\% energy in power-saving mode. Next, we studied that the temperature information is sufficient to infer much useful information regarding an AC for the users - such as energy consumption, the number of compressor cycles - with an accuracy of around 85\%. Lastly, we designed $Greina$ - an online framework to identify the early symptoms of gas leakage in an AC. In our analysis over $90$ outlets, $Greina$ detected 86\% faults at least a week in advance when compared with manual reporting. One must note that, in all the studies, we only sensed room temperature, and gathered climatic conditions from a cloud based weather service, which represents a genuinely low-cost and scalable transition from smart to smarter thermostats. 

\section{Related Work}
	\label{sec:rwork}
Programmable thermostats were one of the initial attempts to save energy in central Heating, Ventilation and Cooling (HVAC) units. Later, Lu et. al.~\cite{lu2010smart} proposed smart thermostat to learn residents' occupancy patterns and control the HVAC. Nest~\cite{Nest} is one such commercial smart thermostat which is available readily available in the market. Studies also evinced the feasibility of energy efficiency in HVAC by using reactive and predictive control of the thermostats based on occupancy~\cite{iyengar2015iprogram, kalaimani2017interaction}. However, one must note that control based on occupants' schedule might lead to inefficient outcomes because tenant's routine is inconsistent~\cite{peffer2011people, kleiminger2014smart}. Besides, studies based on central HVAC is inapplicable to an AC, primarily due to design and operational variations between the two.

Specifically for ACs, manufacturers introduced intelligent and web-based ACs~\cite{Tado, Sensibo} with numerous modes of operations (such as sleep mode) along with smartphone based control. Li et. al.~\cite{li2010optimalc} proposed control algorithm to optimise the duty-cycles of an AC compressor. However, these studies (or products) neither consider the impact of weather conditions, nor analyse the influence of a particular set temperature on tenants' comfort. Furthermore, none of the above mentioned studies monitor the fitness level of AC which is equally critical for the optimised usage of an AC. In independent studies, Ganu et. al.~\cite{ganu2014socketwatch} and Palani et. al.~\cite{palani2014putting} analysed the electricity signature of ACs to find the unusual patterns; but the approach requires user to setup a power meter to monitor AC energy consumption, which is costly and a tedious process than the installation of a smart thermostat. 


\section{System Design}
	\label{sec:studies}
\begin{figure*}[t!]
    \begin{minipage}[t]{0.38\linewidth}
        \centering
		\includegraphics[width=\linewidth]{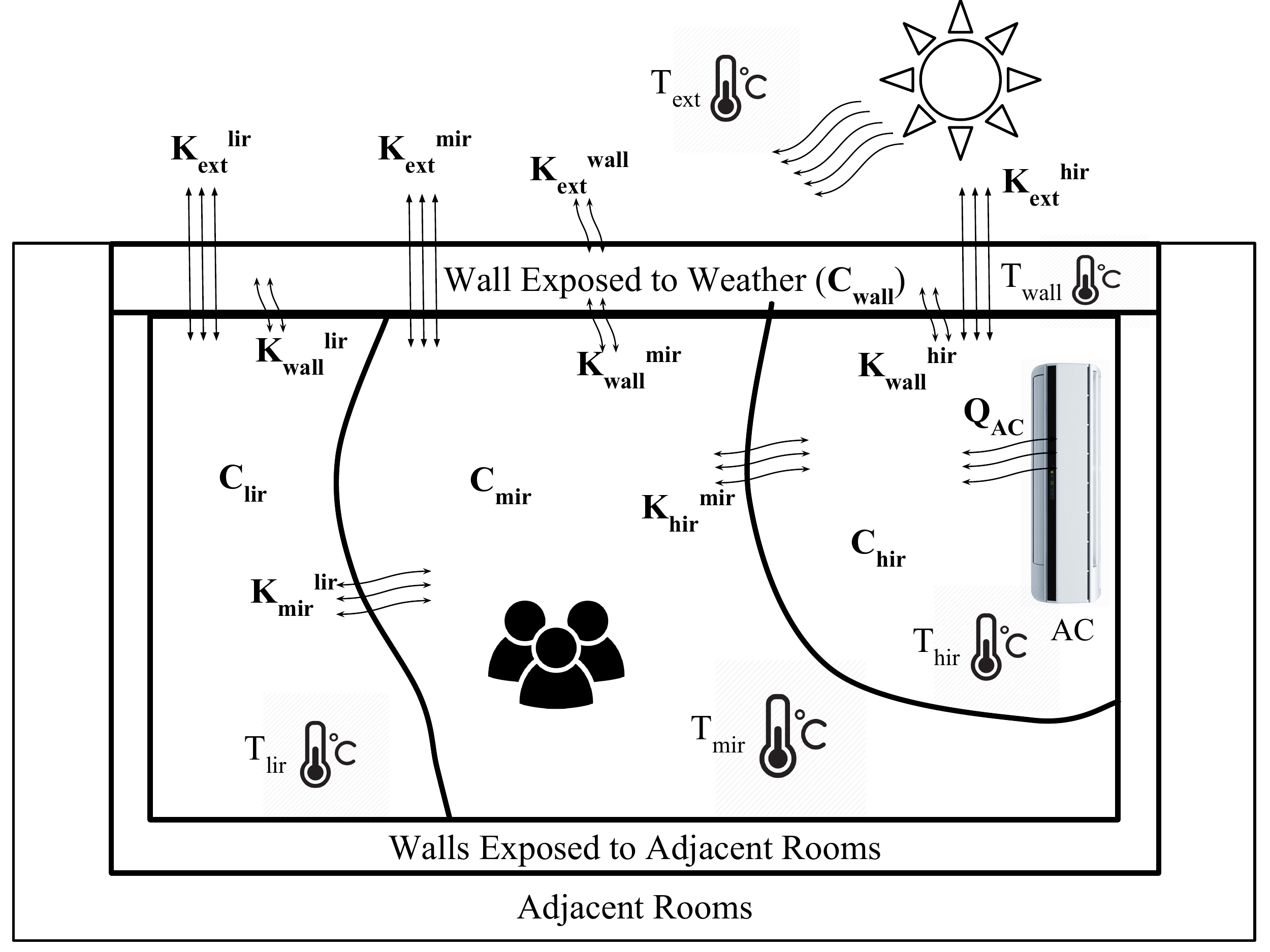}
		\caption{Pictorial representation of various thermal interactions considered occuring in a room divided into three regions.}
		\label{fig:interactions}
	\end{minipage}
    \begin{minipage}[t]{0.28\linewidth}
        \centering
		\includegraphics[width=\linewidth]{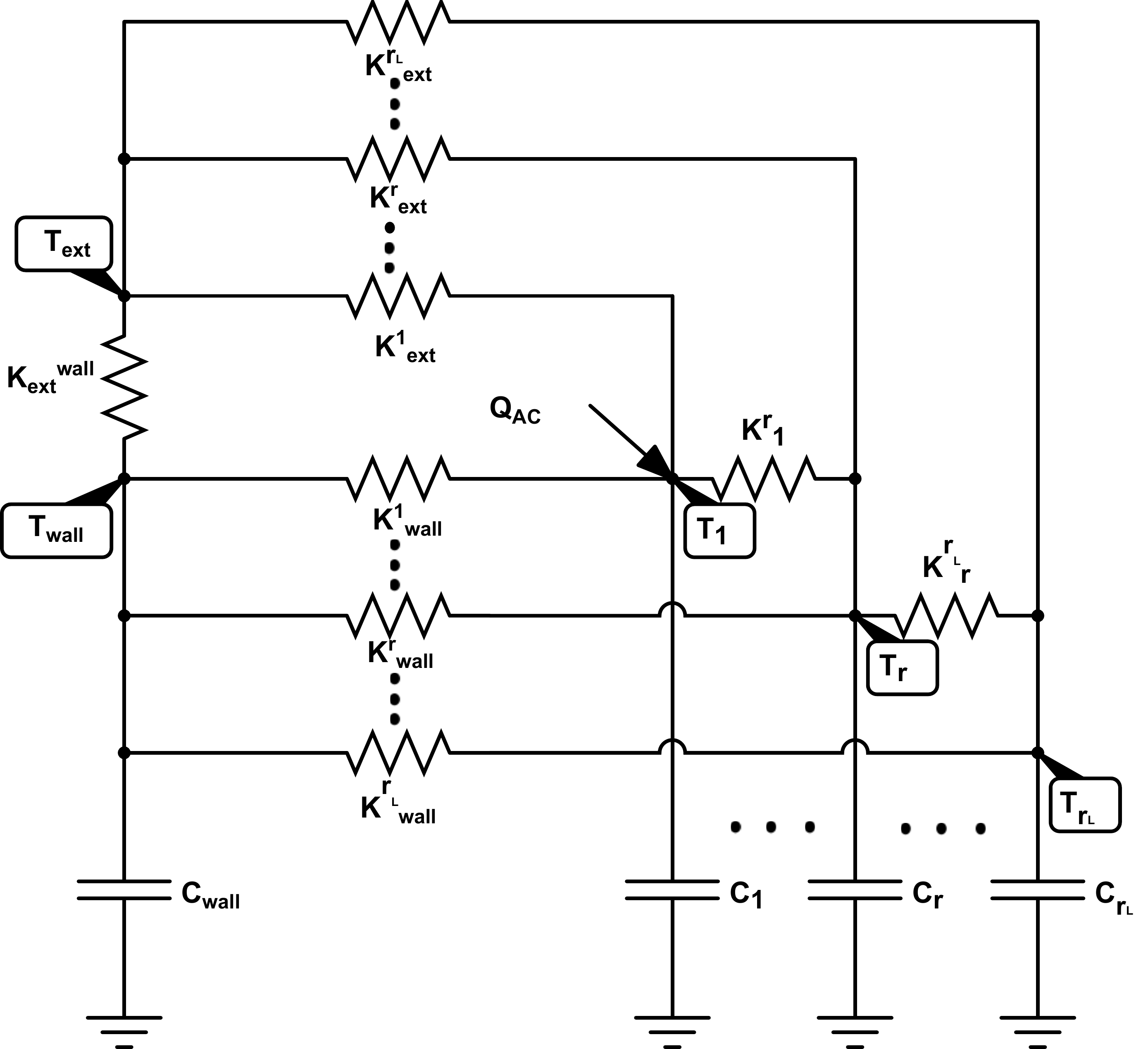}
		\caption{A generic lumped parameter model of a room divided into $n$ zones.}
		\label{fig:rcnetwork}
	\end{minipage}
    \begin{minipage}[t]{0.33\linewidth}
        \centering
		\includegraphics[width=\linewidth]{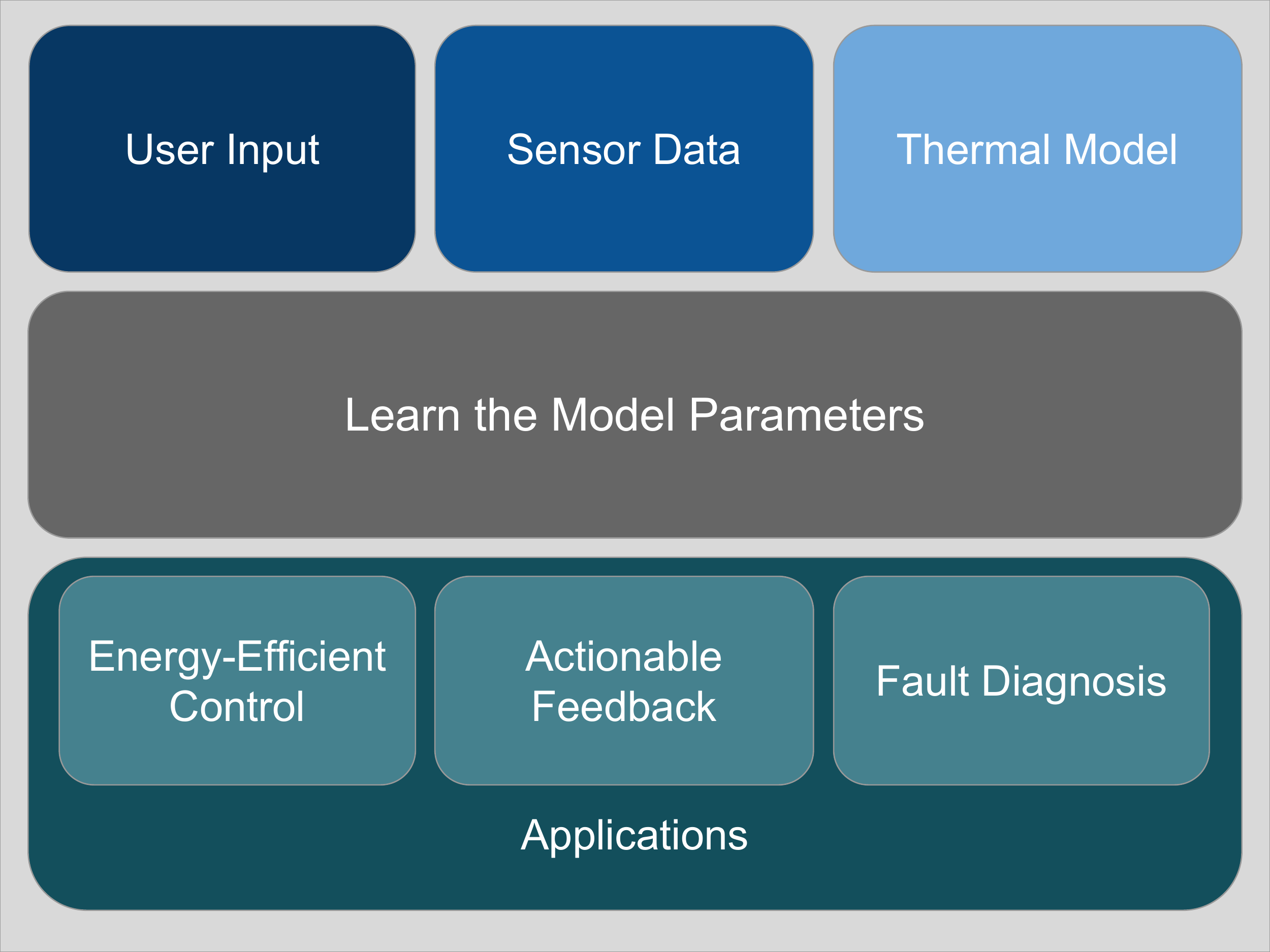}
		\caption{\textbf{System Design:} The input goes to second layer for tuning the model parameters which is then used by the applications.}
		\label{fig:architecture}
	\end{minipage}
\end{figure*}


The thermal behaviour of a space primarily depends on the non-intuitive thermodynamics of the room~(Figure~\ref{fig:interactions}), however, it is hard to monitor each factor affecting the room temperature. Therefore, studies~\cite{madsen1995estimation, radecki2012online} explored \emph{Grey Box Modelling} where sensor data is leveraged to tune the parameters of a lumped thermal model. Figure~\ref{fig:rcnetwork} presents the resistive-capacitive (RC) implementation of one such lumped thermal model of a room divided into $n$ thermal regions. Here, capacitors depict the thermal capacity of a particular area and resistors indicate the heat transfer between any two regions (including the wall and external conditions). The number of parameters of a lumped thermal model is directly proportional to the number of regions in a room ($n$). As tuning involves the actual (temperature) data from a room, the adjusted parameters present an approximate thermal behaviour of the room. 

Given the benefits, we used grey-box model to design the thermostat and the architecture primarly comprises of three layers - 1.) input layer (user, sensor, and model), 2.) learning layer, and 3.) the application layer~(Figure~\ref{fig:architecture})~\cite{jain2017thermalsim}. The thermostat regularly tunes the model parameters to accommodate changes in weather conditions, user activities (in the form of thermal noise), and various other dynamics of the room. The system utilises the tuned thermal model to estimate the room (or region-wise) temperature and the AC compressor state at each time instance. The application layer takes the estimates to then ensure energy-efficient control, actionable feedback for the users, and predictive maintenance of the AC.

\begin{figure*}[t!]
    \begin{minipage}[t]{0.51\linewidth}
        \centering
		\includegraphics[width=\linewidth]{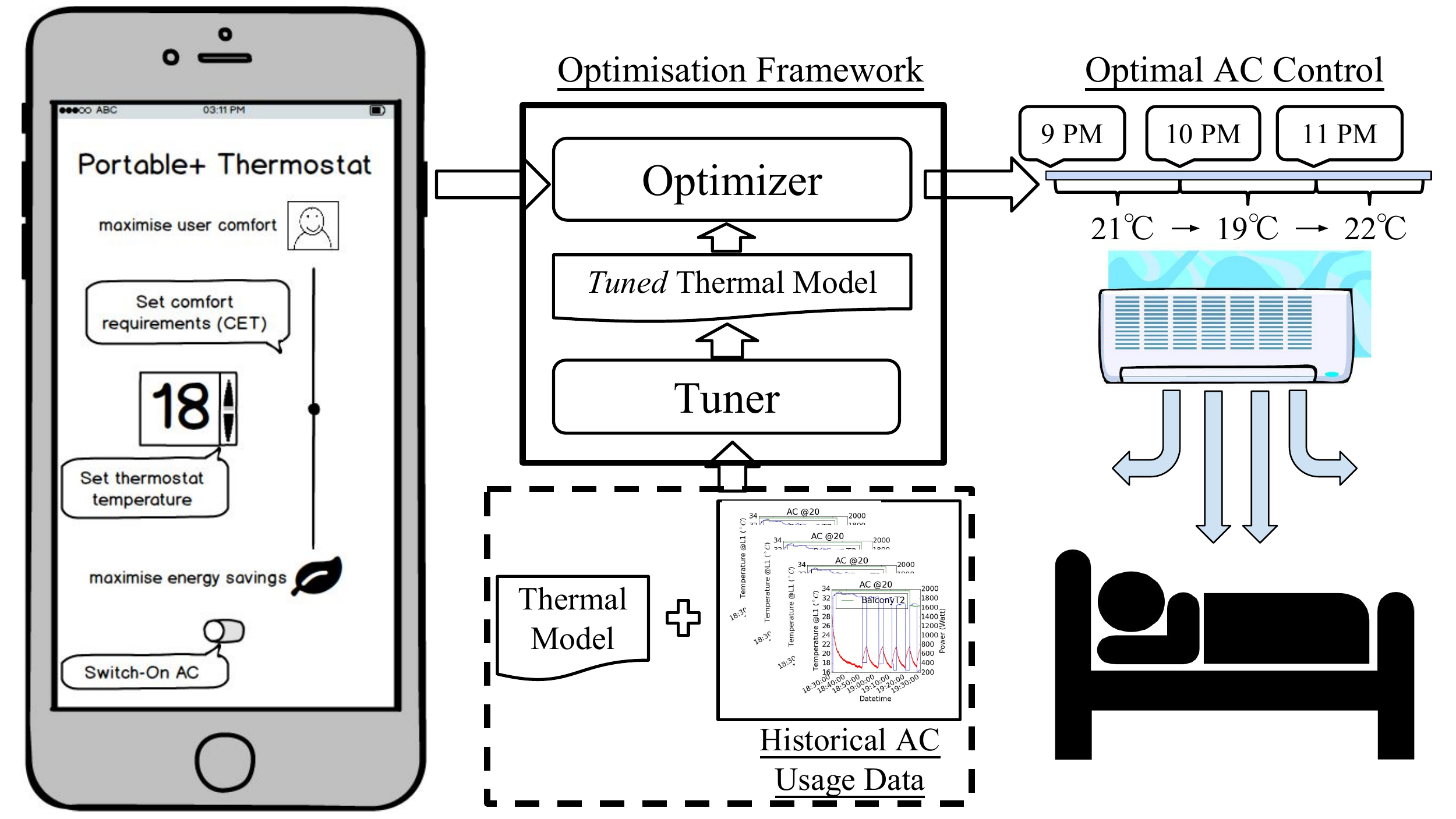}
		\caption{Pictorial representation of the optimisation framework to dynamically change the set temperature based on user input.}
		\label{fig:framework}
	\end{minipage}
    \begin{minipage}[t]{0.49\linewidth}
        \centering
		\includegraphics[width=\linewidth]{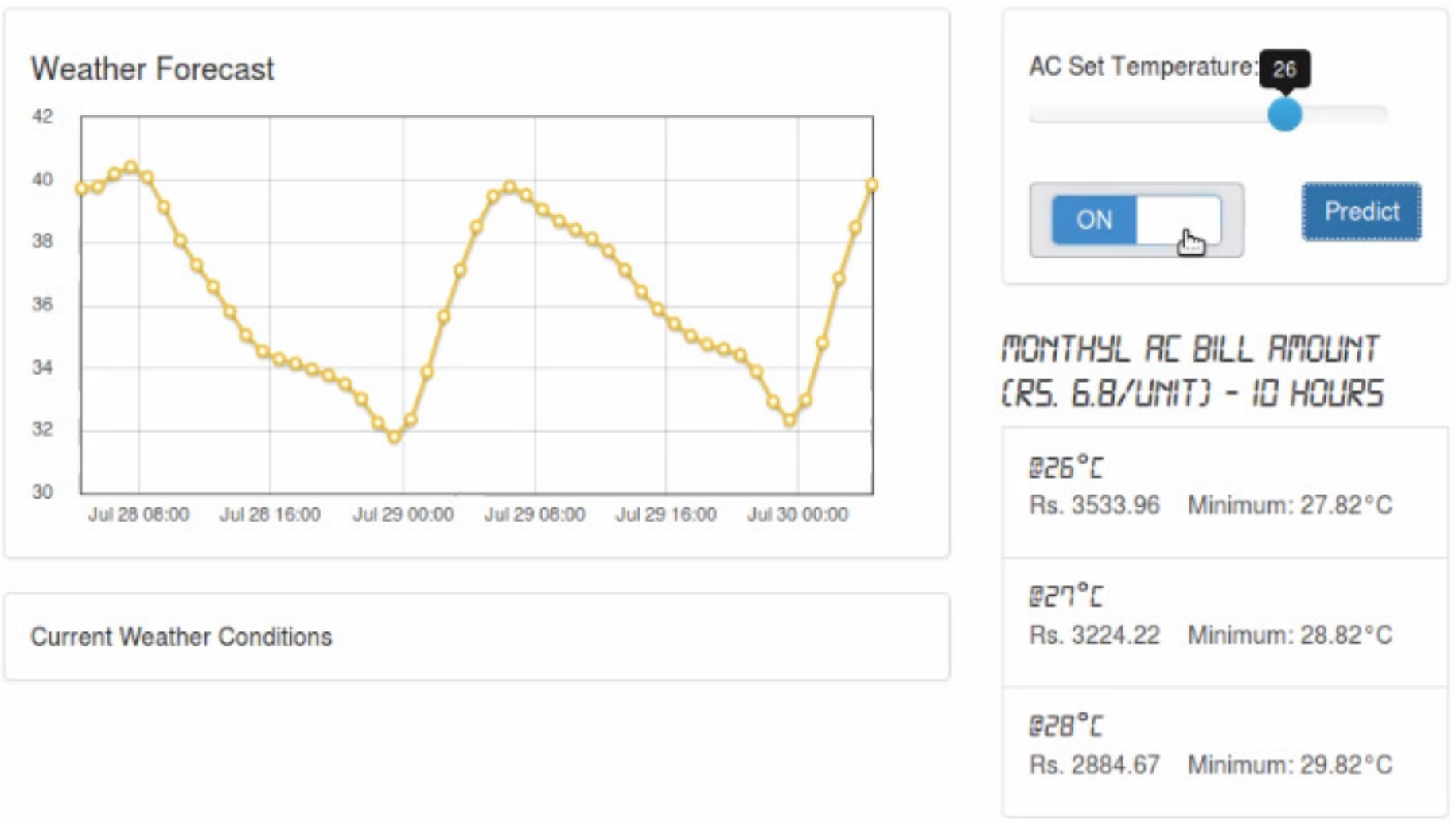}
		\caption{The system takes weather forecast to predict AC energy consumption at different set temperatures for the user.}
		\label{fig:pacman}
	\end{minipage}
\end{figure*}

\subsection{Energy-Efficient Control}
People feel comfortable in a range of temperature. Ideally, a thermostat should report the users, \emph{``What temperature settings will provide them personal comfort and the cost efficiency?''}. We proposed enhancing smart thermostats by adding a Comfort-Energy Trade-off (CET) knob, realised through an optimisation framework which assists users in balancing their comfort and the savings without worrying about the right set temperature~\cite{jain2017portable}. Motivated by Lake Thermal Stratification~\cite{dake1969thermal}, we divided a room into three regions (Figure~\ref{fig:interactions}) - Low Impact Region (\lirWoS), Moderate Impact Region (\mirWoS), and High Impact Region (\hirWoS). 

Here, \hir corresponds to the area in proximity of the AC, facing direct and maximal impact of the cold air coming from the AC. Next, \mir is the region where occupants spend their significant time and often have an indirect effect of AC cooling, while \lir primarily includes the corner spaces of the room. The regions are assumed to be separated by a thin layer of air having negligible thermal mass. The model is an extension of a $2^{nd}$ order thermal model~\cite{dewson1993least} and considers conductive heat transfers due to the difference in temperature of the region (under consideration) and the adjacent spaces such as weather conditions, wall (facing outside), neighbouring regions, and the AC. We assume negligible heat transfer through adjacent rooms within the home. The thermostat (Figure~\ref{fig:framework}) uses the tuned thermal model, set temperature, and CET to then automatically vary the thermostat temperature. Our analysis indicates that the enhancement can reduce occupants' discomfort by 23\% when maximising the user experience, and AC energy consumption by 26\% during the power-saving mode.

\subsection{Actionable Feedback}
Automatically varying the thermostat temperature can sometimes either result in false positives and waste significant energy, or false negatives which can make occupants uncomfortable~\cite{batra2013s}. On contrary, a feedback system allows occupants to override the control logic and specify their preference based on the suggestions. Therefore, we proposed PACMAN - to predict (prior-usage) and estimate (post-usage) AC energy consumption by only sensing the room temperature~(Figure~\ref{fig:pacman})~\cite{jain2014pacman, jain2016non}. 

The advanced algorithm of PACMAN leverages a single-zone thermal model which assumes uniform temperature across the room - a single value indicates the room temperature. A single zone thermal model considers the effect of AC cooling and external temperature while accounting for all other internal and external sources of heat transfer as thermal noise in the room. The learned parameters assist PACMAN in mapping the ambient information (room temperature) with AC energy consumption that otherwise would require plug-level monitoring. The proposed system ensures occupants' participation by providing power consumption feedback to them at successive phases of the AC usage. In our analysis, the proposed system achieved an average accuracy of 85.3\% and 83.7\% in estimating and predicting AC energy consumption, respectively, across all the homes. 

\subsection{Detecting Gas Leakage}
Air conditioners often break down due to ageing and irregular maintenance. A broken AC wastes significant energy and usually fails to maintain desired temperature for the users. While users promptly identify the sudden failures, they usually fail to sense the early symptoms of slow time-varying faults. Refrigerant gas leakage is an example of slow time-varying defect where coolant leaks through the coils (and valves) and leakage slowly diminishes the cooling capacity of the AC. Besides degrading the equipment life, gas leakage results in energy wastage (between 5\%-15\%), never attains the desired temperature, and makes environment hazardous for the living beings~\cite{brambley2005advanced}. To the best of our knowledge, none of the existing commercial thermostats or existing studies (proposing smart thermostats) deal with diagnosing gas leakages. 

We proposed a scalable and low-cost fault detection engine \emph{Greina} - that monitors ambient information (room temperature and door status) to identify the early symptoms of gas leakage. The proposed framework leverages transfer learning to ensure fault detection even when data is inadequate for training. In addition to that, the online algorithm of \emph{Greina} ensures that model adapts to temporal and spatial diversities in the environment. For performance evaluation, we gathered data from $90$ outlets of a retail enterprise for one year. In our analysis, \emph{Greina} detected $86\%$ faults at least a week in advance, when compared with manual reporting. Furthermore, if maintenance team had employed \emph{Greina}, the company could have saved $2x$ energy while minimising the risk to stored food items by keeping the room $5^\circ$C-$10^\circ$C colder every day, when AC had refrigerant leakage.

\section{Discussions}
	\label{sec:discussion}
In this paper, we discussed three major enhancments in smart thermostat which can provide region-specific comfort, actionable feedback, and fault diagnosis by only sensing ambient information from the room. However, one must also note that the current studies neither deals with the thermostat design nor its mobile application. Previous studies show that human-centric prototypes of such devices (and applications) can significantly influence the outcomes. We keep the thread open for the concerned community. 

Similarly, the wide adoption across the community establishes the reason behind the choice of metrics and functions in the existing framework; they are the nut and bolts of the proposed system. Tightening them might boost the performance of designed thermostat and the community is encouraged to study their variants in enhancing the proposed framework. Moreover, climate, users' attitude (towards energy savings), and many other factors differ significantly across the geographies. Though the shown numbers are an indication of better comfort along with notable energy savings, there can be considerable discrepancy across (and within) the countries.

Furthermore, there still exist multiple open research problems for future exploration. The occupants' location (whether they are home/away) can empower the smart thermostats to run model predictive control (MPC) for dynamic set temperature variation in the ACs. Another interesting application exists in small-scale commercial units - such as restaurants and bank branches - which often deploy multiple ACs and possess a tremendous potential for energy savings~\cite{jain2016data, jain2017decision}. The installed ACs typically run on their full capacities without accounting for cooling impact from neighbouring ACs, present in the same room. Though Karmakar et. al.~\cite{karmakar2013coordinated} proposed coordinated scheduling of multiple ACs to reduce the peak power consumption, enhancing the proposed algorithm can also assist in better grid management. 

\section{Conclusion}
	\label{sec:conclusion}
Room-level air conditioners are designed to maintain a suitable temperature for the occupants in relatively small rooms, especially in residential and small-scale commercial buildings. The smart thermostat provides limited energy-saving features but possess significant scope for enhancements. In this paper, we discussed grey-box techniques for smart thermostats to ensure energy-efficient control of the AC, actionable feedback for the occupants, and refrigerant leakage detection to avoid any permanent damage to the appliance. Our analysis over real-world data shows that the proposed enhancements are effective in significantly reducing the AC energy consumption while ensuring the user comfort. In addition to energy-efficient control, the proposed thermostat is powerful enough to provide actionable and corrective feedback to the user with an accuracy better or comparable to state of the art techniques.

%
%
%
%
%

\bibliographystyle{SIGCHI-Reference-Format}
\bibliography{jain18savvy} 

\balance{}

\end{document}